\documentclass[twocolumn,aps,prb,superscriptaddress,floatfix]{revtex4-1}

\usepackage{graphicx}
\usepackage{dcolumn}
\usepackage{bm}
\usepackage{xfrac}

\begin{document}
%

\title{Magnetic excitations in hole-doped Sr$_2$IrO$_4$: A comparison \\ with electron-doped cuprates}

\author{J. P. Clancy}
\affiliation{Department of Physics, University of Toronto, Toronto, Ontario, M5S 1A7, Canada}

\author{H. Gretarsson}
\affiliation{Department of Physics, University of Toronto, Toronto, Ontario, M5S 1A7, Canada}

\author{M. H. Upton}
\affiliation{Advanced Photon Source, Argonne National Laboratory, Argonne, Illinois 60439, USA}

\author{Jungho Kim}
\affiliation{Advanced Photon Source, Argonne National Laboratory, Argonne, Illinois 60439, USA}

\author{G. Cao}
\affiliation{Department of Physics, University of Colorado at Boulder, Boulder, Colorado 80309, USA}

\author{Young-June Kim}
\affiliation{Department of Physics, University of Toronto, Toronto, Ontario, M5S 1A7, Canada}

\begin{abstract}

We have studied the evolution of magnetic and orbital excitations as a function of hole-doping in single crystal samples of Sr$_2$Ir$_{1-x}$Rh$_{x}$O$_{4}$ (0.07 $\le$ x $\le$ 0.42) using high resolution Ir L$_3$-edge resonant inelastic x-ray scattering (RIXS).  Within the antiferromagnetically ordered region of the phase diagram (x $\le$ 0.17) we observe highly dispersive magnon and spin-orbit exciton modes.  Interestingly, both the magnon gap energy and the magnon bandwidth appear to increase as a function of doping, resulting in a hardening of the magnon mode with increasing hole doping.  As a result, the observed spin dynamics of hole-doped iridates more closely resemble those of the electron-doped, rather than hole-doped, cuprates.  Within the paramagnetic region of the phase diagram (0.17 $\le$ x $\le$ 0.42) the low-lying magnon mode disappears, and we find no evidence of spin fluctuations in this regime. In addition, we observe that the orbital excitations become essentially dispersionless in the paramagnetic phase, indicating that magnetic order plays a crucial role in the propagation of the spin-orbit exciton.
\end{abstract}


\maketitle

\section{Introduction}
The physics of spin-orbit-driven oxides, such as the 5d osmates and iridates, has recently attracted intense interest.  These materials display a variety of novel electronic and magnetic ground states, including spin-orbital Mott insulators, spin liquids, topological insulators, and topological (or Weyl) semimetals \cite{Rau_ARCMP_2016, Witczak_ARCMP_2014}.  The layered perovskite iridates Sr$_2$IrO$_4$ and Ba$_2$IrO$_4$ represent one of the most extensively studied families of spin-orbit-driven materials \cite{Huang_JSSC_1994, Crawford_PRB_1994, Cao_PRB_1998, Kim_PRL_2008, Kim_Science_2009, Jackeli_PRL_2009, Kim_PRL_2012, Wang_PRL_2011, Watanabe_PRL_2013, Yang_PRB_2014, Meng_PRL_2014, Hampel_PRB_2015, Kim_Science_2014, Kim_NP_2016, Zhao_NP_2016, Jeong_NC_2017, Zhou_PRX_2017, Klein_JPCM_2008, Qi_PRB_2012, Lee_PRB_2012, Clancy_PRB_2014, Cao_NC_2016, Ye_PRB_2015, Chikara_PRB_2015, Sohn_SR_2016, Chikara_PRB_2017, Louat_PRB_2018, Lupascu_PRL_2014, Kim_NC_2014, Vale_PRB_2015, Calder_PRB_2018, Igarashi_PRB_2014, Okabe_PRB_2011, Boseggia_PRL_2013, Okabe_PRB_2013, Gretarsson_PRL_2016, Liu_PRB_2016, Pincini_PRB_2017}.  These compounds display striking similarities to the high T$_C$ cuprate superconductors, including (1) a variant of the same K$_2$NiF$_4$ crystal structure \cite{Huang_JSSC_1994, Crawford_PRB_1994}, (2) an antiferromagnetic Mott insulating parent compound \cite{Cao_PRB_1998, Kim_PRL_2008, Kim_Science_2009}, (3) quantum spins ($j_{\mathrm{eff}}$ = 1/2), and (4) large magnetic interactions which are well-described by an effectively isotropic Heisenberg exchange Hamiltonian \cite{Jackeli_PRL_2009, Kim_PRL_2012}.  These similarities have inspired many theoretical proposals that superconductivity may be induced in Sr$_2$IrO$_4$ via chemical doping \cite{Wang_PRL_2011, Watanabe_PRL_2013, Yang_PRB_2014, Meng_PRL_2014, Hampel_PRB_2015}.  While early theoretical studies identified electron-doping as a promising route to achieve d-wave superconductivity \cite{Wang_PRL_2011, Watanabe_PRL_2013}, more recent work has also raised the possibility of s- or p-wave superconductivity on the hole-doped side of the phase diagram \cite{Yang_PRB_2014, Meng_PRL_2014}.  However, experimental progress in the search for iridate superconductivity has been slow. No evidence of bulk superconductivity has been found thus far, although there have been several experimental observations reminiscent of cuprate phenomenology, such as the development of Fermi arcs and a d-wave gap in surface-doped Sr$_2$IrO$_4$ \cite{Kim_Science_2014,Kim_NP_2016}. Most recently, reports of odd-parity hidden order in pure and doped Sr$_2$IrO$_4$ \cite{Zhao_NP_2016, Jeong_NC_2017, Zhou_PRX_2017}, potentially compatible with loop-current order, have led to renewed interest in this family of materials.

Among hole-doped iridates, Sr$_2$Ir$_{1-x}$Rh$_x$O$_4$ has been the most thoroughly investigated to date \cite{Klein_JPCM_2008, Qi_PRB_2012, Lee_PRB_2012, Clancy_PRB_2014, Cao_NC_2016, Ye_PRB_2015, Chikara_PRB_2015, Sohn_SR_2016, Chikara_PRB_2017, Louat_PRB_2018, Zhao_NP_2016, Jeong_NC_2017}, thanks to the availability of high quality single crystal samples over a broad range of dopant concentrations.  Replacing Ir with Rh is a somewhat surprising choice for hole-doping.  However, x-ray absorption spectroscopy \cite{Clancy_PRB_2014} and ARPES \cite{Cao_NC_2016} measurements have confirmed that Rh dopant ions in this system preferentially adopt a Rh$^{3+}$ (4d$^6$, S = 0), rather than Rh$^{4+}$ (4d$^5$, S = 1/2), oxidation state, resulting in effective hole-doping.  Although recent x-ray absorption measurements suggest that the oxidation state of the dopant ions may become more complex at higher concentrations \cite{Chikara_PRB_2017}, all studies agree that the hole-doped picture is valid at the low Rh concentrations (0 $\le$ x $\le$ 0.24) of relevance here \cite{Clancy_PRB_2014, Sohn_SR_2016, Chikara_PRB_2017}.

\begin{figure}
\includegraphics{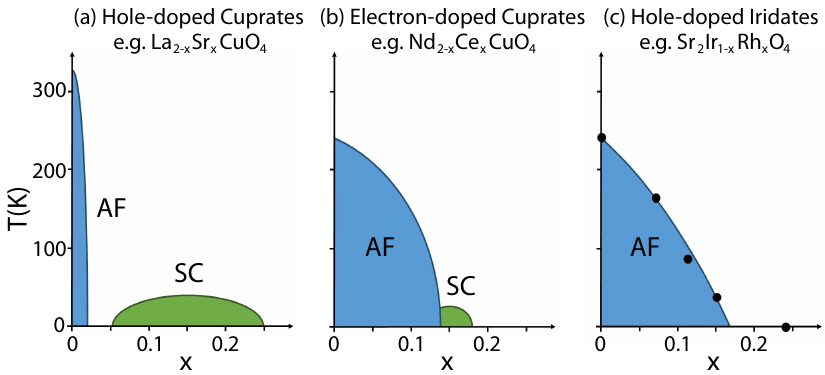}
\caption{ (Color online) Schematic phase diagrams for (a) hole-doped cuprates, (b) electron-doped cuprates, and (c) hole-doped iridates.  Note the strong similarities between the magnetic phase diagrams of the electron-doped cuprates (e.g. Nd$_{2-x}$Ce$_{x}$CuO$_{4}$) and hole-doped iridates (e.g. Sr$_2$Ir$_{1-x}$Rh$_{x}$O$_{4}$).}\label{fig:phase}
\end{figure}

Sr$_2$Ir$_{1-x}$Rh$_x$O$_4$ has a rich electronic and magnetic phase diagram \cite{Qi_PRB_2012}.  Resonant magnetic x-ray scattering measurements indicate a doping-induced change in magnetic structure at x $\leq$ 0.07, with a new canted {\it ab}-plane antiferromagnetic ground state emerging \cite{Clancy_PRB_2014}.  Recent ARPES measurements on lightly doped samples ($x \sim 0.15$) reveal the development of Fermi arcs and a pseudogap reminiscent of the doped cuprates \cite{Cao_NC_2016}.  Above the critical concentration of $x_c$ $\sim$ 0.17, antiferromagnetic order disappears and Sr$_2$Ir$_{1-x}$Rh$_x$O$_4$ becomes a paramagnetic metal/semiconductor.  At higher dopings, more complex magnetic (0.24 $\le$ x $\le$ 0.85) and strongly correlated paramagnetic (x $>$ 0.85) phases have also been observed \cite{Qi_PRB_2012}.  In spite of the absence of superconductivity, we note that the magnetic phase diagram of Sr$_2$Ir$_{1-x}$Rh$_x$O$_4$ is strikingly similar to that of electron-doped cuprates such as Nd$_{2-x}$Ce$_x$CuO$_4$ \cite{Luke_PRB_1990}.  A comparison of these phase diagrams is provided in Fig.~\ref{fig:phase}.

In this paper, we investigate the evolution of the magnetic excitation spectrum in hole-doped iridates by performing high resolution Ir L$_3$-edge resonant inelastic x-ray scattering (RIXS) measurements on single crystal samples of Sr$_2$Ir$_{1-x}$Rh$_x$O$_4$ (0.07 $\le$ x $\le$ 0.42).  RIXS has emerged as the preeminent experimental technique for studying collective excitations in iridates \cite{Kim_PRL_2012, Kim_NC_2014, Lupascu_PRL_2014, Gretarsson_PRL_2013, Gretarsson_PRB_2013, Kim_PRL_2012b, Hozoi_PRB_2014, Gretarsson_PRL_2016, Liu_PRB_2016, Pincini_PRB_2017, Yuan_PRB_2017, Cao_PRB_2017}, providing a sensitive, momentum-resolved probe of spin, orbital, charge, and lattice excitations.  Our measurements on Sr$_2$Ir$_{1-x}$Rh$_x$O$_4$ reveal dispersive spin wave excitations for 0.07 $\le$ x $\le$ 0.15, which appear to {\it harden} as a function of increasing doping.  This provides evidence of yet another intriguing parallel between doped iridates and doped cuprates, as similar magnon hardening has also been reported for electron-doped Pr$_{0.88}$LaCe$_{0.12}$CuO$_{4-\delta}$ \cite{Wilson_PRB_2006} and Nd$_{2-x}$Ce$_x$CuO$_4$ \cite{Jia_NC_2014, Ishii_NC_2014, Lee_NP_2014}.  This similarity in spin dynamics is consistent with the reversal of electron-hole asymmetry first predicted by Wang and Senthil \cite{Wang_PRL_2011}. In fact, recent Ir L$_3$-edge RIXS measurements on the electron-doped iridate Sr$_{2-x}$La$_x$IrO$_4$ \cite{Liu_PRB_2016, Gretarsson_PRL_2016, Pincini_PRB_2017} have revealed more conventional magnon softening, which is reminiscent of the hole-doped cuprates.  This observation lends further support for the reversal of electron-hole asymmetry between doped iridates and cuprates. However, our study also unveils an important difference between the spin excitation spectra in these two families of materials. We find no evidence of short-lived magnetic excitations or paramagnons above x$_c$, indicating that spin fluctuations in the paramagnetic phase are negligible. This observation is also corroborated by our study of orbital excitations. Below x$_c$ we observe two branches of strongly dispersive orbital excitations: those corresponding to transitions between the $j_{\mathrm{eff}}$ = 3/2 and $j_{\mathrm{eff}}$ = 1/2 levels (commonly referred to as the spin-orbit exciton mode) and those corresponding to transitions between the $t_{2g}$ and $e_g$ levels. Above x$_c$ these two branches become essentially dispersionless, highlighting the importance of magnetic order in the propagation of these excitations. These results will be discussed in comparison with doped cuprates as well as electron-doped iridates.

\section{Experimental Details}
Single crystal samples of Sr$_2$Ir$_{1-x}$Rh$_x$O$_4$ ($\sim$1.0 $\times$ 1.0 $\times$ 0.1 mm$^3$ in size) were synthesized using self-flux techniques, as described elsewhere \cite{Cao_PRB_1998, Qi_PRB_2012}.  High resolution Ir L$_3$-edge RIXS measurements (E$_i$ = 11.217 keV) were performed using the MERIX spectrometer on Beamline 30-ID-B at the Advanced Photon Source.  A double-bounce diamond-(1,1,1) primary monochromator, channel-cut Si-(8,4,4) secondary monochromator, and spherical (2 m radius) diced Si-(8,4,4) analyzer crystal were used to obtain an overall energy resolution of 35 meV (FWHM).  Samples were mounted in a closed-cycle cryostat and measured at T = 10 K (with the exception of the x=0.42 sample, which was measured at room temperature).  To minimize the elastic scattering contribution, measurements were carried out in horizontal scattering geometry with a scattering angle close to $2\theta$ = 90$^{\circ}$.  All spectra have been normalized to incident flux, but no additional corrections have been performed.

\section{Experimental Results}

\subsection{Doping Dependence}
\begin{figure}
\includegraphics{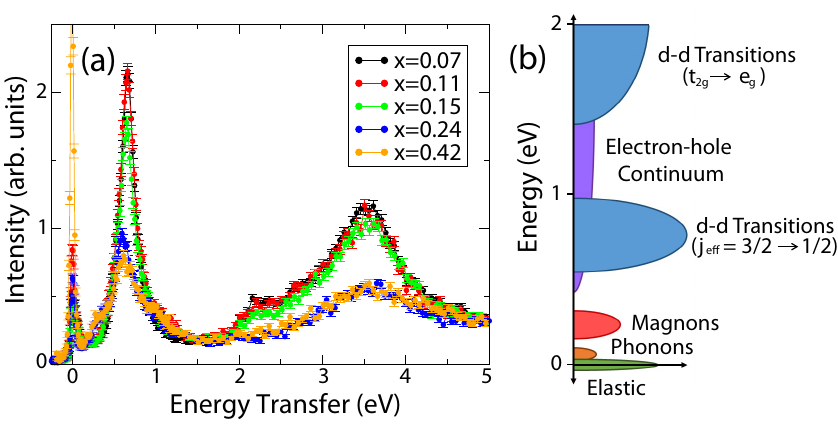}
\caption{ (Color online) (a) Representative RIXS spectra for Sr$_2$Ir$_{1-x}$Rh$_{x}$O$_4$ with 0.07 $\le$ x $\le$ 0.42.  These spectra were collected at {\bf Q} = (0, 0, 33), the magnetic zone center.  (b) A cartoon description of the elementary excitations and energy scales observed in this material.}\label{fig:representative}
\end{figure}

Representative RIXS spectra for Sr$_2$Ir$_{1-x}$Rh$_x$O$_4$ (0.07 $\le$ x $\le$ 0.42) are presented in Fig.~\ref{fig:representative}(a). These energy scans were collected at the {\bf Q} = (0, 0, 33) position in reciprocal space, which corresponds to the center of the magnetic Brillouin zone.  Each spectra contains several distinctive features, including (1) a sharp, resolution-limited elastic line ($\Delta$E = 0), (2) low-lying magnetic scattering (0 $\le$ $\Delta$E $\le$ 0.2 eV), and (3) strong d-d excitations at $\Delta$E $\sim$ 0.7 eV ($j_{\mathrm{eff}}$ = 3/2 to 1/2) and $\Delta$E $\sim$ 3.5 eV ($t_{2g}$ to $e_g$).  In addition to these prominent features, there are also weaker contributions due to phonons (which are essentially negligible at T = 10 K) and particle-hole excitations across the insulating gap (observable as broad continuum scattering above $\Delta$E = 0.4 eV). The heirarchy of excitations in Sr$_2$Ir$_{1-x}$Rh$_x$O$_4$ is illustrated in Fig.~\ref{fig:representative}(b).

\begin{figure}
\includegraphics{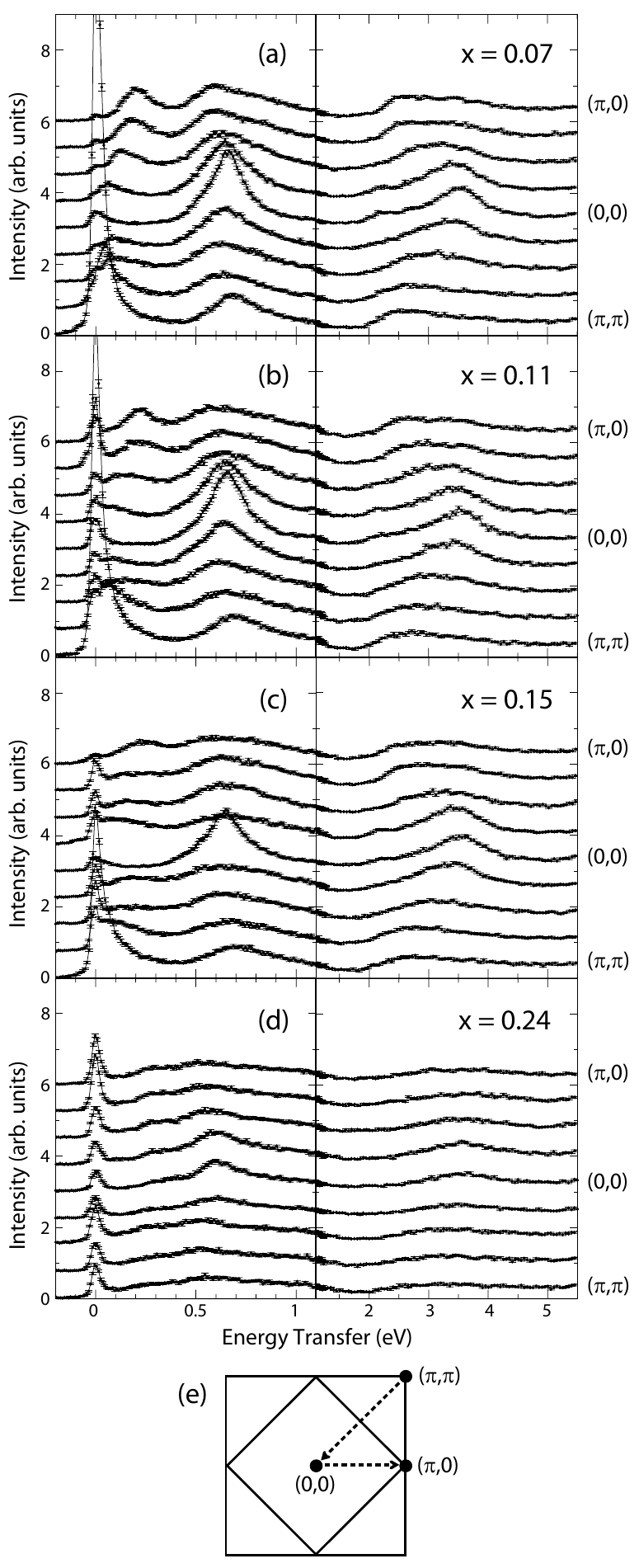}
\caption{ (Color online) Dispersion of magnetic and orbital excitations in Sr$_2$Ir$_{1-x}$Rh$_{x}$O$_{4}$ for x = 0.07 (a), 0.11 (b), 0.15 (c), and 0.24 (d).  Samples with 0.07 $\le$ x $\le$ 0.15 lie within the antiferromagnetically ordered region of the phase diagram, while x = 0.24 is paramagnetic.  Energy scans collected along the line from {\bf Q} = ($\pi$, $\pi$) $\rightarrow$ (0, 0) $\rightarrow$ ($\pi$, 0) have been vertically offset for clarity.  A drawing of the two-dimensional magnetic Brillouin zone is provided in (e).}\label{fig:linescans}
\end{figure}

Note that the spectra presented in Fig.~\ref{fig:representative}(a) display significant doping dependence.  In particular, there is a dramatic difference between spectra collected within the antiferromagnetically ordered (x $<$ 0.17) and paramagnetic (x $>$ 0.17) regions of the phase diagram.  This can be examined in greater detail in Fig.~\ref{fig:linescans}, which illustrates the momentum dependence of RIXS spectra collected for samples with x = 0.07, 0.11, 0.15, and 0.24.  For x = 0.07 to 0.15, we observe dispersive magnon and spin-orbit exciton modes which closely resemble those of the undoped parent compound \cite{Kim_PRL_2012, Kim_NC_2014}.  The magnon mode has a bandwidth of $\sim$200 meV, reaching its maximum energy at the ($\pi$, 0) magnetic zone boundary (i.e. {\bf Q} = (1/2, 1/2, 33)).  The magnon lineshape is significantly broader in the doped compounds, with inelastic peak widths approximately four (x = 0.07) to eight (x = 0.15) times broader than the experimental resolution limit.  The spin-orbit exciton mode exhibits dispersion which is similar in magnitude, but opposite in direction, to the magnon.  At higher energies, we find that the $t_{2g}$ $\rightarrow$ $e_g$ excitations display clear dispersion and strong {\bf Q}-dependent scattering intensity.  In contrast, the x = 0.24 data provides no evidence of low-lying magnetic excitations.  The d-d excitations are much weaker in intensity, and their dispersion is significantly reduced.  Although there still appears to be some weak spectral weight at $\Delta$E $\sim$ 0.3 eV, the data in Fig.~\ref{fig:linescans}(d) show that this feature displays weak momentum dependence which is opposite to that of the magnon mode (i.e. it reaches an energy minimum at the zone boundaries, and an energy maximum at the (0,0) zone center).  As such, it seems unlikely that this feature is magnetic in origin.  A more likely explanation may arise from the presence of small concentrations of Ir$^{5+}$ introduced at higher doping levels \cite{Clancy_PRB_2014, Chikara_PRB_2017}.  The $\sim$0.3 eV energy scale is very similar to that of the low lying d-d excitations previously observed in Ir$^{5+}$ based iridates, such as the double perovskites Sr$_2$MIrO$_6$ (M = Y, Gd) \cite{Yuan_PRB_2017}.

\begin{figure}
\includegraphics{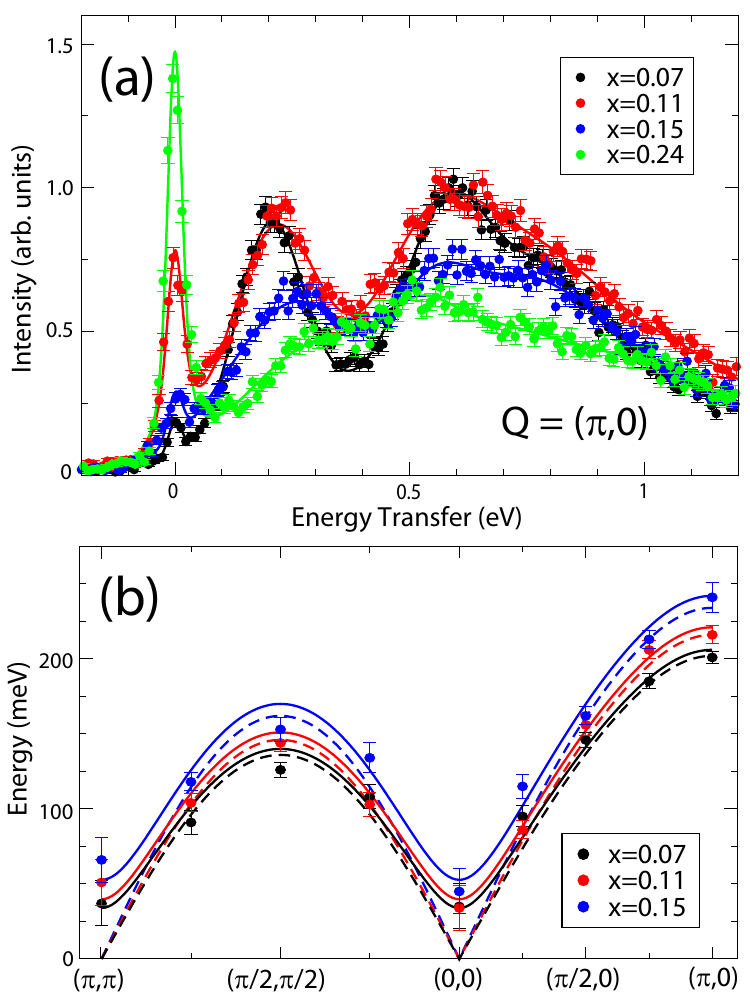}
\caption{ (Color online) (a) Doping dependence of RIXS spectra collected at the ($\pi$, 0) magnetic zone boundary.  As the concentration of holes increases, the position of the low-lying magnon peak shifts towards higher energy.  (b) Dispersion of the magnon mode for x = 0.07 to 0.15.  The experimental dispersion has been fit to an anisotropic Heisenberg model which features gapless in-plane magnon modes (dashed line) and gapped out-of-plane modes (solid line). Details of the fitting and modeling procedure are described in the main text.}\label{fig:dispersion}
\end{figure}

\begin{figure}
\includegraphics{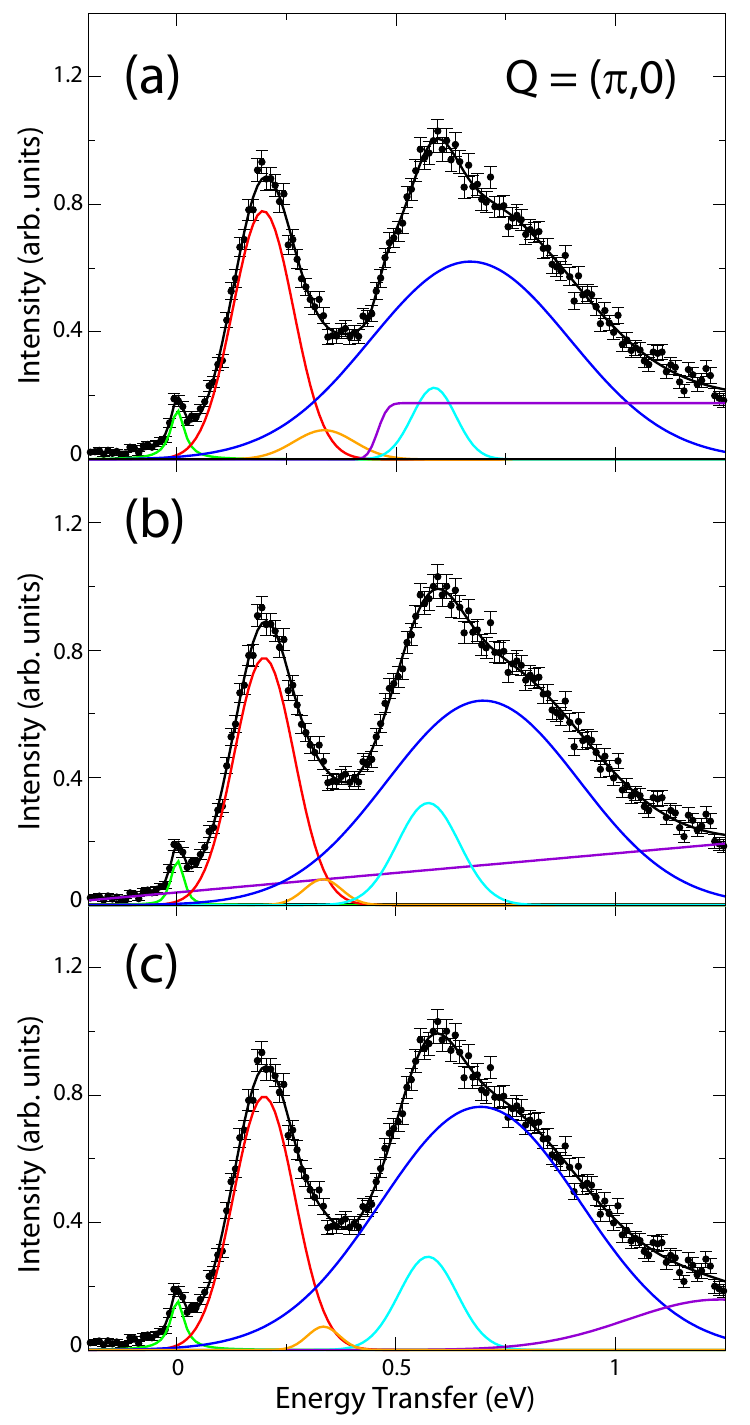}
\caption{(Color online) Representative fits performed for RIXS spectra collected at {\bf Q} = ($\pi$,0) and T = 10 K for Sr$_2$Ir$_{0.93}$Rh$_{0.07}$O$_4$.  Fit components correspond to the elastic line (green), single magnon (red), bimagnon (orange), and $j_{\mathrm{eff}}$ = 3/2 $\rightarrow$ 1/2 orbital excitations (cyan, blue).  Background and electron-hole continuum excitations (purple) have been modeled by (a) a step function, (b) a sloping linear background, and (c) a broad Gaussian peak. }\label{fig:fitting}
\end{figure}

The doping dependence of RIXS spectra collected at the ($\pi$, 0) magnetic zone boundary is illustrated in Fig.~\ref{fig:dispersion}(a).  As the concentration of Rh dopant ions increases, we observe that the magnon broadens and gradually shifts towards higher energy. In order to obtain the quantitative dispersion relation for the magnons, as shown in Fig.~\ref{fig:dispersion}(b), each RIXS spectra was analyzed using a 6-component fit function consisting of elastic line (resolution-limited pseudo-Voigt peak), single magnon (Gaussian), bimagnon (Gaussian), $j_{\mathrm{eff}}$ = 3/2 $\rightarrow$ 1/2 orbital excitations (two Gaussians), and electron-hole continuum (smooth step function). A representative fit carried out using this 6 component function is provided in Fig.~\ref{fig:fitting}(a). To test the robustness of our fitting results, a number of variations on this fit function were also employed.  The inelastic features were modeled using Gaussian, Lorentzian, and mixed combinations of Gaussian and Lorentzian lineshapes.  These approaches revealed negligible differences in peak position (i.e. no variation greater than the experimental uncertainties), and only slight ($\sim$5-10\%) differences in goodness-of-fit.  The most significant source of uncertainty was found to arise from the modeling of the broad electron-hole continuum scattering.  In Fig.~\ref{fig:fitting}(b) and (c) we demonstrate the effect of varying the form of the function used to describe the background/continuum.  Although we observe clear systematic changes in the fit components associated with the orbital and bimagnon scattering contributions, we find that the choice of fit function has very little impact on the single magnon peak position ($\pm$5 meV or less).  This provides a strong indication of the robustness and reliability of the magnon dispersion data in Fig.~\ref{fig:dispersion}(b).

Comparing the magnon dispersion of Sr$_2$Ir$_{1-x}$Rh$_x$O$_4$ to those of the undoped parent compound \cite{Kim_PRL_2012}, we find that at small doping levels (x $\sim$ 0.07) there is very little change in the magnetic dispersion.  This implies that the doping-induced change in canted antiferromagnetic structure between x = 0 (net moments stacked in an A-B-B-A sequence) and x = 0.07 (net moments stacked in an A-A-A-A sequence) has negligible effect on the spin dynamics, a result which reflects the strong quasi-two-dimensional magnetic character of Sr$_2$IrO$_4$.  As the doping increases from x = 0.07 to x = 0.15 we observe a steady increase in magnon energies across the entire Brillouin zone. This observation appears somewhat counterintuitive, since one would expect a {\em softening} of the magnon mode to occur as doping tends to {\em weaken} magnetic order and hence reduce the magnetic interaction strength. In the next subsection, we carry out quantitative dispersion analysis in order to examine the apparent hardening of the magnon dispersion in more detail.

\subsection{Modeling of the Magnon Dispersion}

The magnon dispersion data provided in Fig.~\ref{fig:dispersion}(b) was analyzed using an anisotropic 2D Heisenberg Hamiltonian with nearest neighbor ($J$), second nearest neighbor ($J_2$), and third nearest neighbor ($J_3$) magnetic exchange interactions:
\begin{eqnarray}
{\cal H} &=& \sum_{\langle i,j \rangle} J \left[ S_i^x S_j^x + S_i^y S_j^y + \left(1 - \alpha_{XY} \right) S_i^z S_j^z\right] \nonumber \\
&& + \sum_{\langle \langle i,j \rangle \rangle} J_2 \vec{S}_i \cdot \vec{S}_j + \sum_{\langle \langle \langle i,j \rangle \rangle \rangle} J_3 \vec{S}_i \cdot \vec{S}_j,
 \end{eqnarray}
where $\alpha_{XY}$ is a phenomenological easy-plane XY anisotropy term ($0 \leq \alpha_{XY} \leq 1$). This Hamiltonian was applied in the isotropic limit ($\alpha_{XY}=0$) to fit the first RIXS data on Sr$_2$IrO$_4$ by Kim {\it et al} \cite{Kim_PRL_2012}. However, follow-up high-resolution RIXS measurements by Kim {\it et al} \cite{Kim_NC_2014} suggest the presence of a small ($\sim$ 30 meV) gap in the magnetic excitation spectrum, which motivated the inclusion of XY anisotropy in subsequent studies by Igarashi and Nagao \cite{Igarashi_PRB_2014}, Vale {\it et al} \cite{Vale_PRB_2015}, and Pincini {\it et al} \cite{Pincini_PRB_2017}.  The introduction of XY anisotropy breaks the degeneracy of the magnetic modes associated with in-plane and out-of-plane spin fluctuations.  The in-plane magnon mode remains gapless, while the out-of-plane mode develops a spin gap at the magnetic zone center.

In order to disentangle the doping dependence of $J$ and $\alpha_{XY}$, we have examined our data in both the isotropic limit (Model 1), where we force $\alpha_{XY}=0$, as well as the more general anisotropic case (Model 2), in which $\alpha_{XY}$ is treated as a free fitting parameter. The out-of-plane mode remains gapless in Model~1, while it acquires a gap ($\Delta_\perp$) in Model~2. The in-plane mode remains gapless ($\Delta_\parallel =0$) in both models.
It should be noted that these models are purely two-dimensional, and neglect any dispersion along the $c$-direction due to inter-plane magnetic exchange couplings, which are far too small to be resolved with current RIXS experimental resolution.  Similarly, this model does not include cyclic exchange coupling, J$_C$, which cannot be distinguished from the effects of the ferromagnetic $J_2$.  Recent inelastic neutron scattering measurements on pure Sr$_2$IrO$_4$ report a small in-plane gap of $\sim 0.6$~meV \cite{Calder_PRB_2018}, which also falls beyond the current experimental energy resolution of RIXS.

\begin{figure}
\includegraphics{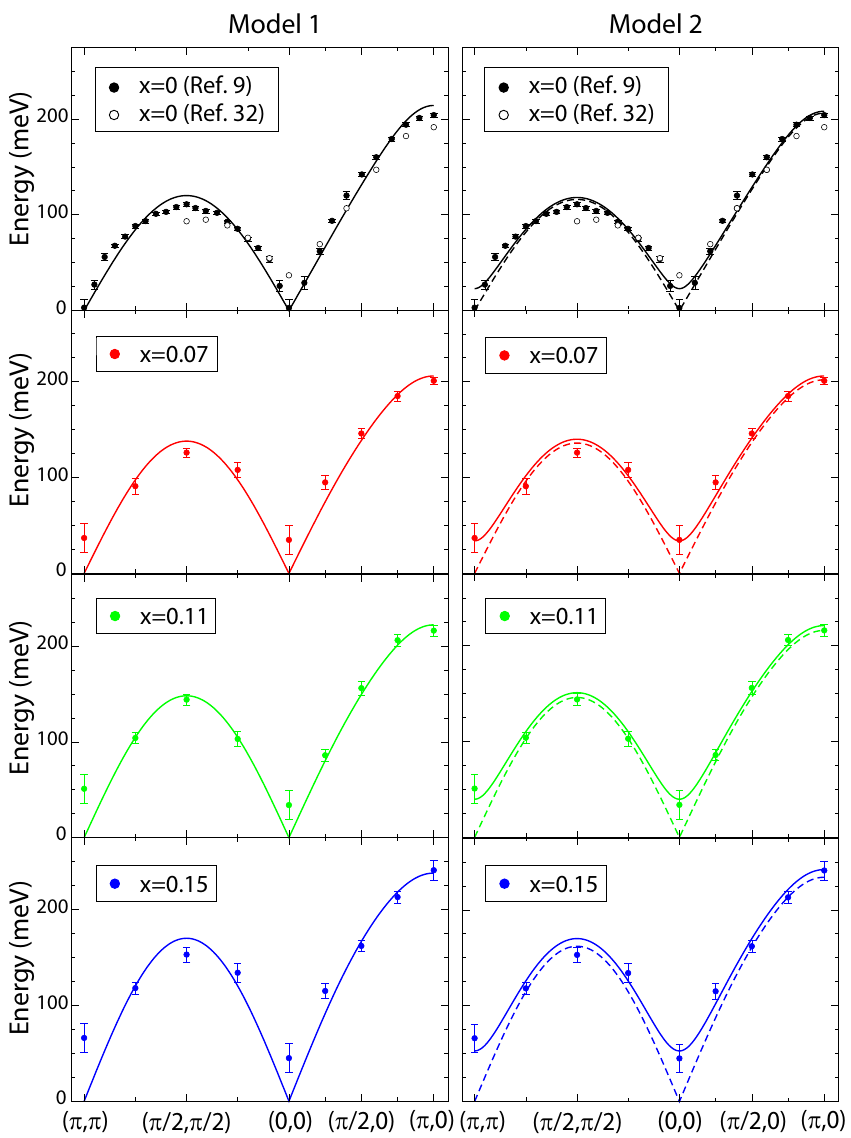}
\caption{(Color online) Magnon dispersion of Sr$_2$Ir$_{1-x}$Rh$_x$O$_4$ for (a) x = 0, (b) x = 0.07, (c) x = 0.11, and (d) x = 0.15.  The dispersion data has been fit using two different theoretical models: Model 1 (left) is an isotropic Heisenberg model (as employed in Refs. \onlinecite{Kim_PRL_2012, Gretarsson_PRL_2016, Liu_PRB_2016}), while Model 2 (right) represents an anisotropic Heisenberg model with easy-plane XY anisotropy (as employed in Refs. \onlinecite{Vale_PRB_2015, Pincini_PRB_2017}).  Magnon dispersion data for x = 0 has been reproduced from Refs. \onlinecite{Kim_PRL_2012} and \onlinecite{Vale_PRB_2015}.}\label{sfig2}
\end{figure}

Figure~\ref{sfig2} shows the magnon dispersion of Sr$_2$Ir$_{1-x}$Rh$_x$O$_4$ as fit to Model 1 (left) and Model 2 (right) for x = 0, 0.07, 0.11, and 0.15.  The magnetic exchange parameters extracted from these two models are provided in Tables I and II.  In all four cases we find that the goodness-of-fit is significantly better for the anisotropic Heisenberg model.  For the undoped parent compound (x = 0) we have reproduced the low resolution (E$_{res}$ = 130 meV) magnetic dispersion data reported by Kim {\it et al.} in Ref. \onlinecite{Kim_PRL_2012}, and the high resolution (E$_{res}$ = 30 meV) data reported by Vale {\it et al.} in Ref. \onlinecite{Vale_PRB_2015} (although based on a re-analysis of measurements performed by Kim {\it et al.} in Ref. \onlinecite{Kim_NC_2014}).  Due to the small energy shift between these two data sets, and the difference in total {\bf Q}-range covered, we have analyzed the x = 0 dispersion in terms of one combined data set.

\begin{table}
\caption{Doping dependence of magnon dispersion parameters from Model 1, the isotropic Heisenberg model (fixed $\alpha_{XY}=0$).  $J$, $J_2$, and $J_3$ represent the first, second, and third nearest-neighbor magnetic exchange interactions.}
\setlength{\tabcolsep}{8pt}
\begin{tabular}{c c c c} 
\hline\hline 
\rule{0pt}{2.5ex}
Doping								& $J$ (meV)				& $J_2$ (meV)			& $J_3$ (meV) 		\\  
\hline 
\rule{0pt}{2.5ex}
x = 0	  			             & 62(5)					& -19(2)				&	13(2)					\\
x = 0.07    					& 67(5)					& -18(2) 				& 8(2)					\\
x = 0.11    					& 77(5)					& -17(2) 				& 10(2)					\\
x = 0.15    					& 87(6)					& -16(3) 				& 9(2)					\\
\hline\hline 
\end{tabular}
\label{table1}
\end{table}

\begin{table*}
\caption{Doping dependence of the magnon dispersion parameters from Model 2, the anisotropic Heisenberg model with $\alpha_{XY}$ as a fitting parameter. $J$, $J_2$, and $J_3$ represent the first, second, and third nearest-neighbor magnetic exchange interactions. The magnitude of the out-of-plane spin gap at the magnetic zone center ($\Delta_\perp$), which is not a fitting parameter, is also included for comparison.  The full Hamiltonian for this model is provided in Eq. (1).}
\setlength{\tabcolsep}{15pt}
\begin{tabular}{c c c c c c} 
\hline\hline 
\rule{0pt}{2.5ex}
Doping										& ${J}$ (meV)		& $J_2$ (meV)		& $J_3$ (meV) 	& $\alpha_{XY}$	& $\Delta_\perp$ (meV)	\\  
\hline 
\rule{0pt}{2.5ex}
x = 0						& 65(5)								& -19(2)				&	13(2)					&	0.02(1)							&	23(5)							\\
x = 0.07    							& 71(5)								& -15(2) 				& 9(2)					& 0.04(1)							&	34(5)							\\
x = 0.11    							& 78(5)								& -15(2) 				& 10(2)					& 0.05(1) 						&	40(5)							\\
x = 0.15    							& 85(6)								& -16(3) 				& 10(2)					& 0.08(1)							&	53(6)							\\
\hline\hline 
\end{tabular}
\label{table2}
\end{table*}

\begin{figure}
\includegraphics{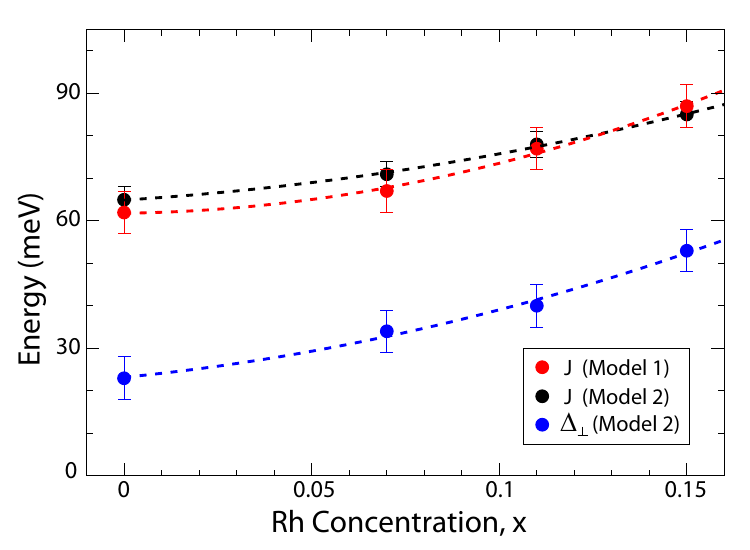}
\caption{(Color online) Doping dependence of several key parameters from the modeling of the magnon dispersion data in Figure \ref{sfig2}.  J is the nearest-neighbor magnetic exchange coupling from Model 1 (red) and Model 2 (black), respectively, while $\Delta_\perp$ is the magnitude of the out-of-plane spin-wave gap at the magnetic zone center (Model 2).}\label{sfig3}
\end{figure}

The key findings from our modeling of the magnetic dispersion are as follows: (1) the nearest-neighbor magnetic exchange interaction ($J$) steadily increases as a function of Rh concentration, regardless of which model is chosen, and (2) when XY anisotropy is included, the magnitude of the out-of-plane spin gap ($\Delta_\perp$) also grows larger as a function of doping.  The doping dependence of the key parameters from Models 1 and 2 are illustrated in Fig.~\ref{sfig3}. Note that the increase in $\Delta_\perp$ alone cannot account for the hardening of the zone boundary energy, and our analysis strongly suggests that {\em both} $J$ and $\Delta_\perp$ grow larger with increasing x. Our analysis indicates that the magnetic bandwidth of Sr$_2$Ir$_{1-x}$Rh$_x$O$_4$ increases by $\sim$10-15\% from x = 0.07 to 0.15, while the nearest-neighbor magnetic exchange coupling increases by $\sim$20-30\%.

\begin{figure}
\begin{center}
\includegraphics{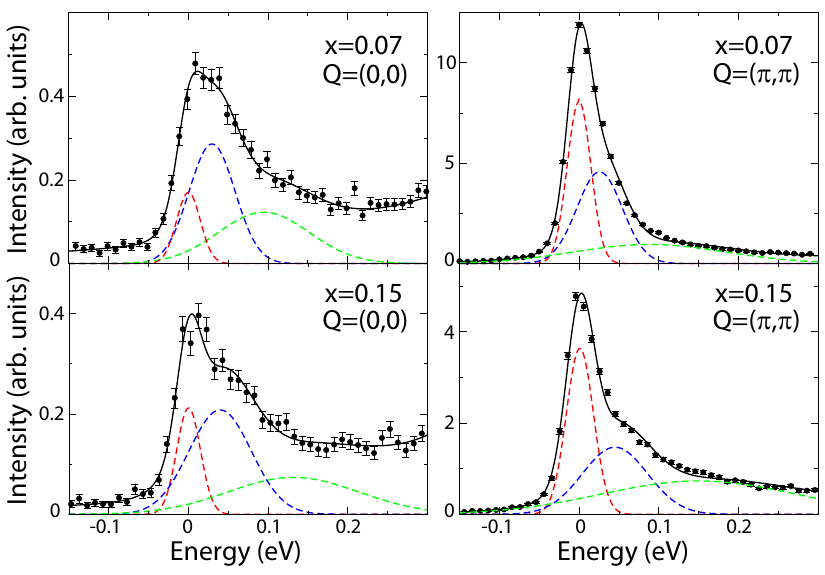}
\end{center}
\caption{(Color online) Representative RIXS spectra collected at the magnetic zone center, (0,0), and magnetic Bragg peak position, ($\pi$,$\pi$), for Sr$_2$Ir$_{1-x}$Rh$_{x}$O$_4$ with x = 0.07 and x = 0.15.  Fit components illustrate the scattering contributions from the elastic line (red), single magnon (blue), and bimagnon (green) excitations.  Note that fitting analysis suggests the presence of a spin gap in both samples, and indicates that the magnitude of the gap increases with doping.}\label{fig:gap}
\end{figure}

Direct evidence for the presence of the out-of-plane spin gap is examined in Fig.~\ref{fig:gap}.  This figure shows representative RIXS spectra collected at the magnetic zone center, {\bf Q} = (0,0), and the magnetic Bragg peak position, {\bf Q} = ($\pi$,$\pi$).  These two {\bf Q} points represent the energy minima for both the in-plane and out-of-plane magnon modes.  Due to the limitations of the current experimental energy resolution, we note that it is not possible to resolve a well-defined spin gap in Sr$_2$Ir$_{1-x}$Rh$_x$O$_4$.  However, we can observe a clear low energy feature on the shoulder of the elastic line, which arises from the single magnon excitation.  Fitting analysis suggests that the minimum energy for this magnon excitation is greater than 30 meV for both x = 0.07 and x = 0.15.  In addition, this single magnon peak clearly shifts towards higher energies at both {\bf Q} positions for x = 0.15, consistent with a spin gap that increases as a function of hole-doping.

In the preceding discussion we have assumed that pure and Rh-doped Sr$_2$IrO$_4$ can be accurately described by a strong-coupling approach based on $J_{eff}=1/2$ local moments.  We note that the limitations of this approach for the iridates have been highlighted in several recent theoretical studies \cite{Arita_PRL_2012, Parschke_NC_2017, Mohapatra_PRB_2017, Stamokostas_PRB_2018}. In particular, there is evidence of strong hybridization between the $J_{eff}=1/2$ and $J_{eff}=3/2$ states \cite{Mohapatra_PRB_2017} as well as the $t_{2g}$ and $e_g$ states \cite{Stamokostas_PRB_2018} due to the substantial itinerancy of these materials. This suggests that the Hubbard model, rather than the Heisenberg model, may provide a better starting point to describe the magnetic excitations in our system.  In this case, we have chosen to use the anisotropic Heisenberg model because we are primarily interested in providing a phenomenological analysis of the gap size and the overall bandwidth of the magnetic excitations.  Despite these limitations, we note that the Heisenberg model is often used to describe magnetic excitations in other itinerant systems (e.g. iron pnictides). 

\subsection{Dispersion of Orbital Excitations}

\begin{figure}
\begin{center}
\includegraphics{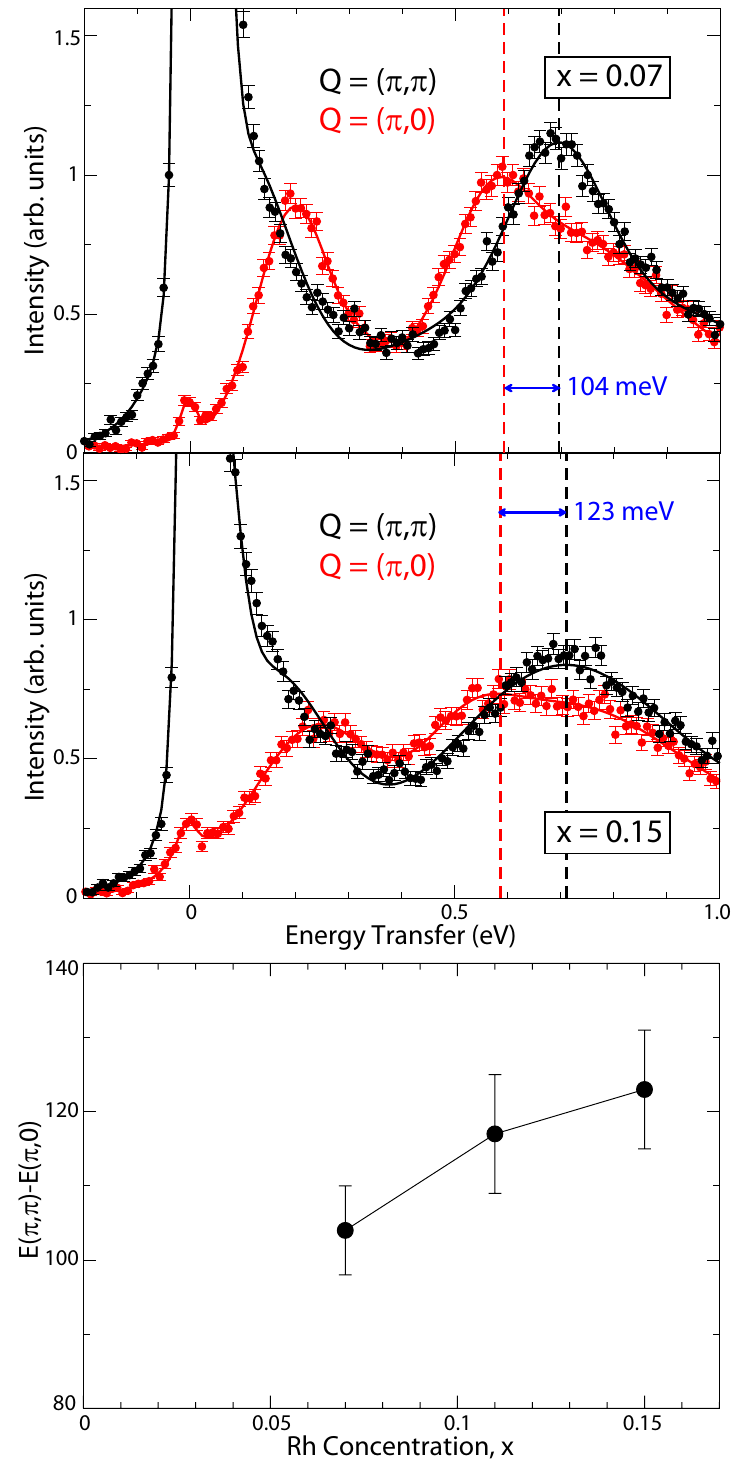}
\end{center}
\caption{(Color online) Dispersion of orbital excitations in Sr$_2$Ir$_{1-x}$Rh$_x$O$_4$ for (a) x = 0.07 and (b) x = 0.15.  Representative RIXS spectra collected at {\bf Q} = ($\pi$,0) and ($\pi$,$\pi$) provide a measure of the orbital bandwidth.  The doping dependence of this bandwidth, E($\pi$,$\pi$)-E($\pi$,0), is illustrated in panel (c).}\label{fig:orbiton}
\end{figure}

The dispersion of the orbital excitations in Sr$_2$IrO$_4$ is intimately coupled to the dispersion of the magnetic excitations.  As a result, one might expect that a significant hardening of the magnetic excitations should also be reflected in the energy of the orbital excitations.  The spin-orbit exciton mode is a highly dispersive orbital excitation which arises from  d-d transitions between the $j_{\mathrm{eff}}$ = 3/2 and $j_{\mathrm{eff}}$ = 1/2 states. Propagation of this excitation involves spin-flips costing energy on the scale of $\sim 2J$. Analyzing the dispersion of the spin-orbit exciton mode allows us to confirm the observed magnon hardening effects, while avoiding potential complications arising from the limited experimental energy resolution, the strong elastic scattering contributions at {\bf Q} = (0,0) and ($\pi$,$\pi$), and the doping-dependent spin gap.  However, modeling of the orbital dispersion is complicated by the fact that the $j_{\mathrm{eff}}$ = 3/2 $\rightarrow$ 1/2 excitations are split by a non-cubic crystal electric field, overlap with the electron-hole continuum scattering, and display strongly {\bf Q}-dependent scattering intensity.  For this reason we focus on the doping dependence of the orbital bandwidth, as measured by the energy difference between the peaks of the orbital excitations observed at {\bf Q} = ($\pi$,0) and ($\pi$,$\pi$).  Note that {\bf Q} = ($\pi$,0) is the energy maximum for the magnon mode, and hence an energy minima for the spin-orbit exciton.  Similarly, {\bf Q} = ($\pi$,$\pi$) is one of two energy maxima (along with {\bf Q} = (0,0)).  In order to ensure consistency, the peak of the orbital excitations was determined both through fitting analysis and through inspection of the raw data.  As shown in Fig. \ref{fig:orbiton}, we observe a small, but significant, increase in the bandwidth of the orbital excitations upon doping.  We note that the $\sim$9\% increase in orbital bandwidth between x = 0.07 and x = 0.15 is fully consistent with the 10\% (Model 2) to 16\% (Model 1) increase observed in the magnon bandwidth.  Thus, we conclude that hole-doped Sr$_2$Ir$_{1-x}$Rh$_{x}$O$_4$ displays hardening of both magnetic and orbital excitations.

\section{Discussion}

Our study of the evolution of magnetic excitations in Sr$_2$Ir$_{1-x}$Rh$_x$O$_4$ as a function of doping allows us to draw two important conclusions.  Firstly, the disappearance of the magnon mode above x$_c$ $\sim$ 0.17 indicates that there are negligible spin fluctuations within the paramagnetic phase of Sr$_2$Ir$_{1-x}$Rh$_x$O$_4$.  This represents a clear distinction from the doped cuprates, where short-lived paramagnon excitations have been observed well into the highly overdoped regime \cite{Dean_NM_2013, LeTacon_NP_2011, LeTacon_PRB_2013}. The suppressed magnetic fluctuations in the paramagnetic phase of hole-doped Sr$_2$IrO$_4$ also contrasts with the observation of persistent paramagnons in electron-doped Sr$_2$IrO$_4$ \cite{Gretarsson_PRL_2016, Pincini_PRB_2017} or persistent (largely dispersionless) antiferromagnetic excitations in Ru-doped Sr$_2$IrO$_4$ \cite{Cao_PRB_2017}. This observation is also corroborated by the reduced dispersion of the d-d excitations. Strongly dispersive orbital excitations are one of the most distinctive features of the Sr$_2$IrO$_4$ RIXS spectrum.  This dispersion arises as a consequence of the strong spin-orbit coupling in this material, which means that as an excited $j_{\mathrm{eff}}$ = 3/2 hole propagates through the antiferromagnetically ordered background it leaves behind a trail of misaligned $j_{\mathrm{eff}}$ isospins \cite{Kim_PRL_2012, Kim_NC_2014}. The lack of dispersion for the orbital excitations at x = 0.24 is therefore consistent with the absence of the magnetic order.

Another surprising observation is the hardening of the magnon mode in the antiferromagnetically ordered phase. A similar form of magnon hardening has also been observed in electron-doped cuprates such as Pr$_{0.88}$LaCe$_{0.12}$CuO$_{4-\delta}$ \cite{Wilson_PRB_2006} and Nd$_{2-x}$Ce$_x$CuO$_4$ \cite{Jia_NC_2014, Ishii_NC_2014, Lee_NP_2014}.  This effect is particularly striking in Nd$_{2-x}$Ce$_x$CuO$_4$, where the magnetic bandwidth effectively doubles from x = 0 to x = 0.15.  Initially, this doping-induced hardening appears to be quite counterintuitive, as one would expect the cost of a spin flip excitation to be reduced by the dilution of the antiferromagnetic background.  Indeed, a doping-induced broadening and softening of the magnon mode is observed in many hole-doped cuprates as well as in electron-doped Sr$_2$IrO$_4$ \cite{Gretarsson_PRL_2016, Liu_PRB_2016, Pincini_PRB_2017}. However, in the case of a locally static hole model, the energy gain from magnetic dilution can be offset by a reduction of hole delocalization energy (i.e. three-site exchange) compared to the undoped system.  This mechanism was originally proposed for electron-doped Nd$_2$CuO$_4$ by Jia {\it et al.} \cite{Jia_NC_2014}.  As in the case of the magnetic phase diagrams (Fig. 1), it appears that the spin dynamics of hole-doped Sr$_2$Ir$_{1-x}$Rh$_x$O$_4$ more closely resemble those of the electron-doped, rather than hole-doped, cuprates. We would like to point out that Rh doping is not exactly comparable to the usual hole- or electron-doping by substitution {\em away} from the CuO$_2$ or IrO$_2$ layers. In addition to affecting the carrier concentration, Rh doping also appears to affect the nature of the Ir and Rh magnetic moments \cite{Clancy_PRB_2014}.

With this caveat in mind, it is still worthwhile to examine the analogy between the iridates and cuprates. The apparent reversal of electron-hole asymmetry between cuprates and iridates was first pointed out by Wang and Senthil \cite{Wang_PRL_2011}, and is expected based on differences in electronic structure (iridates have electron-like Fermi surfaces, while cuprates have hole-like Fermi surfaces).  In particular, this effect is believed to arise from the sign of $t'/t$ (where $t$ and $t'$ represent the nearest-neighbor and next-nearest-neighbor hopping terms, respectively) when the electronic structure is mapped onto an effective one-band Hubbard model.  This quantity is positive for iridates and negative for cuprates.  Thus, even though the magnitude of $t'/t$ is quite similar in both families of materials, the superconducting phase diagrams predicted by the one-band Hubbard model are not.  Indeed, recent theoretical calculations indicate that d-wave superconductivity will only be stable in electron-doped iridates, such as Sr$_{2-x}$La$_x$IrO$_4$ \cite{Watanabe_PRL_2013, Yang_PRB_2014, Meng_PRL_2014}.  However, functional renormalization group calculations predict s-wave superconductivity in hole-doped Sr$_2$IrO$_4$ \cite{Yang_PRB_2014}, and argue that the effects of Hund's rule coupling will reduce (enhance) superconductivity on the electron (hole) doped side of the phase diagram.  In addition, dynamical mean field theory (DMFT) calculations suggest that when Hund's rule coupling is sizable, p-wave superconductivity may arise in the hole-doped iridates \cite{Meng_PRL_2014}.  Unlike the cuprates, Parschke {\it et al} have recently argued that correlation-induced effects will give rise to fundamental differences between the electron and hole-doped sides of the iridate phase diagram, extending well beyond the simple reversal of $t'/t$ [\onlinecite{Parschke_NC_2017}].  It is clear that further systematic doping studies will be required on both electron and hole-doped iridate systems.

\section{Conclusions}

In summary, we have used RIXS to investigate the magnetic and orbital excitations of the hole-doped spin-orbital Mott insulator Sr$_2$Ir$_{1-x}$Rh$_x$O$_4$.  We observe low-lying magnon excitations, which harden with increasing doping, and disappear within the paramagnetic phase.  These results add to the growing list of similarities between hole-doped iridates and electron-doped cuprates, and offer intriguing clues for the ongoing search for iridate superconductivity.  In addition, we observe strongly dispersive orbital excitations ($j_{\mathrm{eff}}$ = 3/2 $\rightarrow$ 1/2 and $t_{2g}$ $\rightarrow$ $e_g$) at low doping, which become essentially dispersionless within the paramagnetic phase.  This illustrates the importance of magnetic order to the dynamics of the spin-orbit exciton mode, and highlights the unique spin-orbital character of this system.  We hope these results will help to guide and inform future work on Sr$_2$IrO$_4$ and other doped iridate materials.

\begin{acknowledgements}
The authors would like to acknowledge valuable discussions with M.P.M. Dean, Y. Cao, and D.S. Dessau.  Work at the University of Toronto was supported by NSERC of Canada.  GC acknowledges support from NSF under Grant No. DMR-1712101.  Use of the Advanced Photon Source at Argonne National Laboratory is supported by the U.S. Department of Energy, Office of Science, Office of Basic Energy Sciences, under Contract No. DE-AC02-06CH11357.
\end{acknowledgements}

\end{document}